\documentclass[aip,jcp,citeautoscript,unsortedaddress]{revtex4-1}

\usepackage{graphicx}%
\usepackage{dcolumn}%
\usepackage{bm}%

\begin{document}


\title{Large electronic bandwidth in solution-processable pyrene crystals: The role of close-packed crystal structure}


\author{Fran\c coise Provencher}

\author{Jean-Fr\'ed\'eric Laprade}

\author{Gabrielle Simard}

\author{Nicolas B\'erub\'e}

\author{Michel C\^ot\'e}

\author{Carlos Silva}
\email[Corresponding author. E-mail address: ]{carlos.silva@umontreal.ca}
\affiliation{D\'{e}partement de physique et Regroupement qu\'{e}b\'{e}cois sur les mat\'{e}riaux de pointe, Universit\'{e} de Montr\'{e}al,  C.P.\ 6128, Succursale centre-ville, Montr\'{e}al (Qu\'{e}bec), H3C~3J7, Canada}

\author{Julien Tant}

\author{V\'eronique de Halleux}

\author{Yves Geerts}
\affiliation{Laboratoire de chimie des polym\`eres, Universit\'e Libre de Bruxelles, Code Postal 206/1, Boulevard du Triomphe, 1050 Bruxelles, Belgium}


\date{\today}

\begin{abstract}
We examine the interdependence of structural and electronic properties of two substituted pyrene crystals by means of combined spectroscopic probes and density-functional theory calculations. One derivative features n-hexyl side groups, while the other one contains branched silanyl groups. Both derivatives form triclinic crystal structures when grown from solution, but the electronic dispersion behavior is significantly different due to differences in $\pi$--$\pi$ overlap along the $a$ crystal axis. Both systems display dispersion of 0.40--0.45\,eV in the valence band, suggesting a high intrinsic hole mobility. However, the dispersion is primarily along the $a$-axis in the silanyl-substituted derivative, but less aligned with this crystal axis in the hexyl-substituted material. This is a direct consequence of the diferences in co-facial $\pi$ electron overlap revealed by the crystallographic studies. We find  that photophysical defects, ascribed to excimer-like states, point to the importance of localized trap states. Substituted pyrenes are useful model systems to unravel the interplay of crystal structure and electronic properties in organic semiconductors.
\end{abstract}


\maketitle



\section{INTRODUCTION\label{Sec:Intro}}

The electronic properties of organic semiconductors incite tremendous scientific interest due to their applications in optoelectronic devices such as photovoltaic and electroluminescent diodes, as well as in field-effect transistors. Self-organised molecular crystals, in which long-range order enhances charge transport and optical properties, are the subject of intensive research activity in the context of such applications. Solution-processable materials favor simple crystal growth techniques over high-temperature vacuum-deposition methods, but aliphatic side groups generally play a central role in determining the crystal structure and hence the electronic structure of organic semiconductor films. In addition, they may contribute substantially to its structural and energetic disorder. For example, vacuum-grown rubrene crystals demonstrate field-effect mobilities as high as 15 -- 20\,cm$^{2}$\,V$^{-1}$s$^{-1}$~\cite{Sundar:2004p3262, Podzorov_PRL_2004}, while the best solution-processed crystals of substituted rubrene derivatives display mobilities of order 0.7\,\,cm$^{2}$\,V$^{-1}$s$^{-1}$~\cite{StingelinStutzmann:2005p3261}. Indeed, on the basis of infrared spectroscopic measurements at room temperature and density-functional theory (DFT) calculations, Li et~al.\ have demonstrated that light quasiparticles with effective mass comparable to the free-electron mass dominate the charge transport in rubrene single-crystal~\cite{Li_PRL_99_2007}. Resonance Coulomb $\pi$--$\pi$ intermolecular interactions are the cornerstone of the electronic properties of organic crystals and are highly dependent on the microscopic molecular packing.  This packing is affected strongly by modifying the molecular side chains~\cite{An_Adv_Mater_17_2005,Ashizawa_Chem_Mater_20_2008} and, as a result, charge mobilities vary significantly upon side-chain substitution. 

In this communication, we compare the crystal and electronic structures of two different pyrene derivatives by means of single-crystal X-ray diffraction, steady-state absorption and photoluminescence (PL) spectroscopy, and density-functional theory (DFT) calculations. The pyrene core is particularly interesting in this context since it offers numerous possibilities of peripheral group modifications, with profound consequences on the condensed-matter structure. For example, unsubstituted pyrene forms monoclinic crystals, while 1,3,6,8-tetraphenyl pyrenes form liquid-crystalline columnar phases~\cite{Hayer_JPCB_2006}, while pyrenes similarly substituted with n-diethyleneglycolether derivatives display a liquid phase at room temperature~\cite{deHalleux_JPPA_2006}. The two derivatives that we study here (Fig.~\ref{fig_struct}) differ by the side chains located on the 1,3,6,8-positions of the pyrene core, with one consisting of a n-hexyl chain (Py-hexyl) while the other is a branched silanyl (Py-silanyl). They both form triclinic crystals, but the side-chains have a strong influence on the detailed molecular arrangement. This is seen from the larger red-shift shift of the Py-silanyl crystal compound compared to the Py-hexyl one -- the needle-like structures of Py-silanyl crystal appear red.  This is a direct consequence of differences in the face-to-face stacking distance of these molecules which maximizes the $\pi$--$\pi$ overlap.  The crystalline spectroscopic properties are discussed in light of band structure calculations for both structures.  We calculate a valence bandwidth of almost 0.5\,eV for Py-silanyl molecular crystal along the $\pi$-stacking axis.  This value is significantly higher than in unsubstituted rubrene crystals.  We find that the contrasting crystal structures, determined by the aliphatic substituents, result in significantly different PL spectra. We attribute the differences in localized excimer-like defects in one crystal structure featuring more co-facial orientation of molecules along a crystal axis. We conclude that in addition to the large electronic dispersion, the possiblity of such photophysical defects is important in crystal structures that feature cofacial packing.

\section{METHODOLOGY\label{Sec:Method}}

\subsection{Experimental Methodology\label{Subsec:Experim}}


X-ray diffraction measurements were performed on single crystals of Py-hexyl and Py-silanyl using a  Bruker Microstar diffrac\-tometer operating with a rotating anode generating CuK$\alpha$ radiation (1.54\,{\AA}). The structure was solved and refined by the laboratory of X-ray diffraction at the Universit\'e de Montr\'eal. \footnote{Crystallographic data (excluding structure factors) for the structures reported in this paper have been deposited with the Cambridge Crystallographic Data Centre as supplementary publication no. CCDC 703228 and CCDC 703229. Copies of the data can be obtained free of charge from www.ccdc.cam.ac.uk/conts/retrieving.html or on application to Cambridge Crystallographic Data Centre (CCDC),
12 Union Road, Cambridge CB2 1EZ, UK.  Tel: (+44) 1223 336 408 Fax: (+44) 1223 336 033
E-mail: deposit@ccdc.cam.ac.uk Web Site : www.ccdc.cam.ac.uk}

The single crystals used for this experiment were slowly grown from solution in the following manner. A small vial containing a concentrated solution (5\,mg/mL) of pyrene derivative Py-hexyl in spectrophotometric grade chloroform (Sigma Aldrich) as the solvent was enclosed in a larger vial containing acetone (Recochem inc.), thus exposing the solution to acetone vapor. Acetone is a poor solvent for Py-hexyl, so its solubility decreased slowly and the molecules then self-assembled into small, needle-shaped crystals. Py-silanyl crystals were slowly grown from an acetone solution (0.3\,mg/mL) and formed long, needle shaped crystals.

Polarized optical micrographs (POM) were acquired by a Nikon D80 SLR camera mounted on a Carl Zeiss microscope equipped with two polarizers. The crystalline films studied by POM were produced by drop-casting a concentrated solution onto a glass microscope slide.

Absorption spectra were acquired with a Varian Cary 500 spectrophotometer. Quartz cuvettes with a 10\,mm pathlength (Hellma) were used to hold the solution. 

The light source used for acquiring  the Raman spectra was a Ti:sapphire laser pumped by a Verdi-V6 laser. The scattered light was harvested by a double spectrometer U-1000 with a focal length of 1\,m and gratings with 1800\,grooves/mm. The spectra were acquired by a photomutiplier tube and a photon counter. The data were spectrally corrected. Raman spectra of Py-hexyl and Py-silanyl in chloroform solution (3\,mg/mL) were obtained for a 300\,mW and 7005\,{\AA} laser beam. 


Profilometry measurements were acquired with a stylus profilometer (Veeco Dektak 150) in map scan mode, using a stylus of radius 12.5 $\mu m$ and applying a stylus force of 1.00 mg.

\subsection{Theoretical Methodology\label{Subsec: Theory}}

The crystalline electronic properties have been computed in the framework of density functional theory by using the Abinit~\cite{Abinit} program.  The exchange and correlation energy was treated with the PBE~\cite{PerdewPBE} functional.  Special care was paid to structural parameters and electronic properties during the convergence study.  The results shown here were obtained with a $3 \times3 \times3$ \emph{k}-point Monkhorst-Pack grid, a cutoff energy of 35 Hartree and a tolerance on the maximum force of $5 \times 10^{-5}$ hartree/Bohr.  
Despite the well-known inability of DFT to describe van der Walls interactions~\cite{Zhao_J_Chem_Theory_Comput_2005}, it has been shown~\cite{Hummer_Phys_Rev_B_2003} that restricting the primitive cell to experimental parameters yields crystalline structures in good agreement with experiment.  We find that the internal parameters of the optimized Py-hexyl crystal fit remarkably the experiment.  Indeed, the distance between two molecular planes, the pitch angle and the roll angle, all fit experimental data within 1.5\%.  In the case of the Py-silanyl crystal, the deviation is more significant.   While the distance between two molecular planes is larger by 5\%, the pitch and roll angles are lower by 3\% and 13\% respectively.  However, the use of the optimized or experimental structures to compute the band diagrams does not affect the shape of the bands. It is worth noting that all intramolecular bonds fit the experimental ones within 2.5\%.  For the interatomic angles, the deviation from experimental measurements is less than 1\%, except in one of the branches whose \emph{sp}-carbons and the silicon atoms show an angle 1.7\% too narrow.

The molecular electronic properties have been obtained with the Gaussian 03~\cite{Gaussian03} package and the B3LYP~\cite{Becke_J_Phys_Chem_98_1993,Lee_Phys_Rev_B_37_1988} functional.  The 6-311G(d) basis set was employed with \emph{Tight} convergence criteria (maximum force converged to 1.5d(-5) Hartree with a residual mean square of 1.0d(-5) Hartree) and a SCF convergence criterion of 1.0d(-10) Hartree.  Time-dependent DFT (TDDFT) calculations allowed us to assess the optical transitions of the molecules.

\section{RESULTS\label{Sec:Results}}

In this work we investigate pyrene cores substituted with two different side chains, which are located on the 1,3,6,8-positions of the molecule. One is a linear hexyl chain (Py-hexyl) while the other is a branched silanyl chain (Py-silanyl), see Fig.~\ref{fig_struct}. These molecules self-assemble in solution such that millimeter-size crystals are readily obtained by solution processing in a poor solvent atmosphere (Fig.~\ref{fig_maille}(a)). Smaller crystals of a few dozen $\mu$m in length can be grown by solution casting. Surface profilometry measurements on several crystals such as those in Fig.~\ref{fig_maille}(a) reveal a height of $3.52 \pm 0.04$\,$\mu$m, indicating a uniform crystal growth. One of the main advantages of these crystals is that their growth is simple, reproducible, and requires no controlled environment such as inert atmosphere, high temperature, or vacuum.

To obtain a detailed understanding of the molecular packing of Py-hexyl and Py-silanyl in crystalline form, we have performed single-crystal X-ray diffraction measurements. These revealed that both pyrene derivatives form triclinic crystal lattices with the unit cell parameters reported in Table~\ref{tab_XRD}. The molecular packing in the Py-hexyl and Py-silanyl crystals are depicted in Fig.~\ref{fig_maille}(b) and Fig.~\ref{fig_maille}(c). The unit cell in both cases contains only one molecule. The distance between two molecular planes is 3.09\,\AA ~and 3.34\,\AA ~for Py-hexyl and Py-silanyl respectively.  To describe the molecular orientation, we use the terminology introduced by Curtis et al.~\cite{Curtis_JACS_2004} which defines the pitch angle as the molecular slipping along the long molecular axis and the roll angle as the molecular slipping along the short axis.  The Py-hexyl crystal is characterized by a pitch angle of 45.8 degrees and a roll angle of 66.9 degrees.  These angles give rise to a molecular slipping of 3.18\,\AA ~and 7.25 \AA ~along the long and short axis respectively.  Since the pyrene core is 7.04 \AA ~long by 4.91 \AA ~wide, there is no direct $\pi$-stacking in the case of the Py-hexyl crystal.  As can be seen in Fig.~\ref{49b_stacking}, the linear chains of adjacent molecules lie just above the conjugated cores.  In the case of the Py-silanyl crystal, the pitch angle is 58.0 degrees and roll angle is 13.0 degrees.  The offset from a face-to-face $\pi$-stack is therefore 5.34\,\AA ~along the long axis and only 0.77\,\AA ~along the short axis, resulting in a direct overlap of the $\pi$ orbitals.

We explore the consequences of these distinct crystal structures on the electronic properties by presenting steady-state absorption and photoluminescence (PL) measurements of dilute solutions of both materials, and in their polycrystalline films (Fig.~\ref{fig_crystal}(a) and~\ref{fig_crystal}(b), respectively). Since the dilute solutions showed no signs of aggregation nor excimeric behavior, we consider that the photophysics of the dilute solutions are representative of those of the isolated  molecule. The vibronic progression of the solution spectra is clearly resolved and is similar for the two chromophores studied. We note that the spectrum of the Py-silanyl molecule is red-shifted by about 60\,meV compared to Py-hexyl and that the shoulder that appears on the 0-0 transition of Py-silanyl is more pronounced. The absorption and fluorescence spectra are mirror images of each other, indicating that the potential energy surface of the ground and excited states have a similar curvature.  The Stokes shift for both molecules is 20\,meV, indicating that the excited-state reorganization energy is small. We have carried out DFT calculations to disect the molecular electronic structure of these pyrene derivatives and the nature of the vibronic progression observed in Fig.~\ref{fig_crystal}(a). The energy level of the HOMO and LUMO orbitals are presented in Table~\ref{energy_molecule}.  These results were obtained for isolated molecules without any solvent effect correction.  The Py-silanyl HOMO and LUMO orbitals can be seen in Fig.~\ref{fig_orbitale}.  We note that they do not differ from those of Py-hexyl corresponding orbitals.  Both wave functions are delocalized over the pyrene core but also expand over the \emph{sp}-hybridized carbon atoms from which stem the side chains.  These atoms therefore strongly affect the electronic properties and are responsible for the reduced HOMO/LUMO energy gap compared to bare pyrene (only hydrogen atoms) which has a theoretical HOMO/LUMO  gap of 3.84\,eV. While orbitals are antisymmetric under a long molecular axis folding, the HOMO and LUMO are respectively symmetric and antisymmetric under short axis folding.  This difference has a major consequence on the crystalline electronic structure as will be discussed later. 

The energy of the first transition peak at the onset of absorption is 2.89\,eV for Py-hexyl and 2.83\,eV for Py-silanyl.  This compares well to the first allowed optical transitions that are computed with the time-dependent DFT (TDDFT) formalism.  In the case of Py-hexyl molecule, we obtained a transition at 2.79\,eV with a 0.908 oscillator strength which is dominated by the HOMO and LUMO orbitals.  For the Py-silanyl molecule, the first transition is also predominantly a HOMO $\rightarrow$ LUMO at 2.71 eV with an oscillator strength of 0.910.  Quantum mechanical calculations of the molecular normal modes that project onto the electronic transition reveal that the observed spectral structure is dominated by a composite progression of five modes, as described in detail in the appendix. These modes have frequencies of 284, 417, 1278, 1380, and 1653\,cm$^{-1}$, with Huang-Rhys parameters ranging from 0.12 to 0.45.

One of the most distinctive features of these results is the 0.74-eV redshift between the solution and crystal PL spectra of Py-silanyl, while those in Py-hexyl redshift by only 0.31\,eV from solution to crystal. The PL spectra of films  of various thicknesses (not shown here), cast from solutions of concentration ranging from 1.7 to 24.4 $\mu$g/cm$^2$, showed  no variation in shape, therefore ruling out the phenomenon of self-absorption.

The absorption spectrum of a Py-hexyl polycrystalline film exhibits clearly resolved vibronic structure, mimicking the molecular absorption spectrum. This is consistent with the absorption spectrum reported for the pure pyrene crystal~\cite{Birks1975}.  On the other hand, the Py-silanyl absorption spectrum is very broad and mostly featureless. The 0-0 transition appears merely as a shoulder at 2.76\,eV, while the vibronic replica remain unresolved.  It also exhibits a long, non-gaussian slope that is usually related to disorder. We show the photoluminescence excitation spectrum of the Py-silanyl sample, which mimics the absorption spectrum, ruling out scattering effects as the source of this tail.

We have calculated the band structure of Py-hexyl and Py-silanyl crystals, which are presented in Fig.~\ref{band_structure}. The symbols of the Brillouin zone high-symmetry points are defined in the caption.  The Py-hexyl crystal, with valence and conduction band edges at the A and $\Gamma$ points ($(\frac{1}{2}\frac{1}{2}0)$ and (000)), respectively, has an indirect bandgap of 1.48\,eV.  However, the direct transition is only 70\,meV higher in energy.  This gap is much smaller than the gap reported for the isolated molecule but this is principally due to the functional that we used for the computation under periodic conditions --- see Section~\ref{Subsec: Theory}. The maximum dispersion of the valence band is 0.44\,eV and is obtained along the path $(\frac{1}{2}\frac{1}{2}0) \rightarrow (0\frac{1}{2}\frac{1}{2})$ (AC) while the conduction band shows its highest dispersion along the $(000) \rightarrow (\frac{1}{2}0\frac{1}{2})$ ($\Gamma$D) axis and amounts to 0.22\,eV.  The Py-silanyl crystal has a direct bandgap of 1.5\,eV.  All directions of the reciprocal lattice that have a component along the $\Gamma$B axis are characterized by a valence band dispersion in the range 0.40 --  0.45\,eV.  This direction corresponds to the short \emph{a}-axis of the primitive cell along which the molecules are $\pi$-stacked. We note that the dispersion  along other directions ($\Gamma$E and $\Gamma$D) is also significant because those vectors display a significant projection onto that in the $\Gamma$B direction. There is no overlap of the molecular orbitals along the other directions as can be seen form the featureless bands in the $\Gamma$F and $\Gamma$G directions, in contrast to the case in the Py-hexyl crystal.  The conduction band of the Py-silanyl crystal is narrower than the valence band with a total width of 0.10\,eV. This is understood by the combination of the LUMO orbital symmetry and the intermolecular arrangement that results in a wave function energy that varies modestly with the phase in \emph{k}-space.  Indeed, higher lying empty bands broaden with, for example, a 0.29\,eV dispersion along $\Gamma$D for the second empty band.

\section{DISCUSSION\label{Sec:Disc}}

Other reports have recognized the key role of the molecular side chains on the solid-state packing~\cite{Ashizawa_Chem_Mater_20_2008,Shin_J_Mater_Chem}.  Here, we demonstrate the same influence and we highlight the fact that side-chain modification can lead to new materials with altered electronic properties. The introduction of main-group elements (such as Si in the silyl groups) to the molecule through side-chain modification is of great interest since it can profoundly affect intermolecular interactions.  Indeed, the intermolecular distance of 3.34 \AA ~that characterize the Py-silanyl crystal is surprisingly short when we consider that Coulomb repulsions render face-to-face configurations energetically unfavorable.  In fact, this distance is less than twice the carbon van der Waals radius ($2 \times 1.70$\,\AA{} = 3.40\,\AA)~\cite{Bondi_J_Phys_Chem_1964}.  Indeed, $\pi$-stacked crystals generally show an intermolecular distance of 3.4 to 3.6 \AA~\cite{Koren2003, Yamamoto1998, Politis2001, Kim_JACS_2007}.  The Py-silanyl molecule is characterized by a branched chain bearing a silicon core atom that has a lower electronegativity than carbon.   Based on a Mulliken population analysis, a charge of 1.22 electrons is present on the silicon atoms.  The electrostatic interaction between this charge and the neighboring positive charges lying on the pyrene core tighten the molecules together and therefore increases the orbital overlap.  This observation opens the door to new developments in the design of organic crystals.  Silicon is the \emph{$sp^3$}-bonding atom with the lowest electronegativity, but it could be interesting to test the effect of using atoms of group IIIB as the core atoms of branched chains.  On a more general basis, the use of main-group elements in organic crystals can be regarded as an efficient way to improve the interatomic overlap of electronic wave functions and, consequently, improve the charge mobility.

Pyrene's optical properties have been studied extensively in the past~\cite{Birks1975}. Compared to the bandgap of pure molecular pyrene in solution (3.33 eV~\cite{Birks1968}), that  of Py-hexyl and Py-silanyl at 2.89\,eV and 2.83\,eV are redshifted. This is explained by the larger extent of the wave function over the the pyrene core and the $sp$ hybridized carbon atoms from which the chains stem. As can be seen from Fig.~\ref{fig_crystal}(a), the shape of the  absorption and emission spectra in solution of the two pyrene derivatives are very similar and thus are independent of the side chains, as was reported earlier for other pyrene derivatives~\cite{halleux04}. Indeed, the HOMO and LUMO, which are the two orbitals implicated in this optical transition as supported by the TDDFT calculations, are localized  on the discotic core and thus are very similar for the two molecules, which explains the similarity observed in the spectra.

We now turn our attention to the PL spectra of polycrystalline films showed in Fig.~\ref{fig_crystal}. Py-hexyl has two PL peaks, one at 2.57\,eV that is narrower than the other at 2.35\,eV. This second peak is reminiscent of the spectral  signature of a spatially extended emitter. We thus attribute the PL spectrum of Py-hexyl to the fluorescence of the first excited singlet state and possibly an excimer-like emitter. In contrast, Py-silanyl PL spectrum is featureless and much lower in energy, two typical features of excimeric emission. We thus attribute the Py-silanyl PL spectrum to an excimer, a species which is known to exist in pure pyrene crystals and cofacial molecular crystals in general~\cite{Birks1975}. In addition, the shift between the PL peak of Py-silanyl in solution (molecular) and in film (excimer) is 740\,meV, in striking accordance with the 6000 cm$^{-1}$ rule characteristic of the emission energy of excimers relative to that of the isolated molecule~\cite{Birks1964}. It is worth noting that this low-energy peak does not stem from a broader peak that would have been sculpted by self-absorption, because we have ruled out this phenomenon by the independence of the PL spectral bandshape on film thickness. 

The highest hole mobility at room temperature in organics has been measured in highly purified pentacene crystal at 35 cm$^2$V$^{-1}$s$^{-1}$ and was clearly attributed to band transport by its temperature dependence~\cite{Jurchescu_APL_2004}.  Different theoretical calculations evaluated the valence bandwidth of pentacene in the range 0.5 -- 0.6\,eV~\cite{de_Wijs_Synth_Met_139_2003,Cornil_Adv_Mater_2001}.  Another well studied organic crystal, rubrene, has a valence bandwidth of 0.34\,eV~\cite{benjamin} and mobility as high as 20 cm$^2$V$^{-1}$s$^{-1}$~\cite{Podzorov_PRL_2004}.  With a valence band dispersion of 0.44 -- 0.45\,eV, the crystals presented here have an electronic structure that would lead to high hole mobility.  It is quite remarkable that, despite the fact that Py-hexyl and Py-silanyl crystals differ considerably in their molecular orientation due to their respective side chains, their band dispersion are comparable.  In the case of Py-hexyl, the large roll angle gives rise to an orbital overlap along a diagonal direction of the lattice and this is reflected in the band structure.  On the other hand, the Py-silanyl molecules have a partial cofacial overlap of their $\pi$ electrons and this $\pi$-stacking direction is in accordance with the maximum dispersion direction of the Brillouin zone.  

\section{CONCLUSIONS\label{Sec:Conclusion}}

We have investigated the electronic structure of two substituted pyrene derivatives, one containing a linear n-hexyl chain and another containing a branched silanyl chain. By means of X-ray diffraction measurements on single crystals, we find that both derivatives display triclinic crystal structures with one molecule in the unit cell, but the $\pi$--$\pi$ overlap along the $a$ axis of the primitive cell is significantly different in the two structures. The inter-core orientation of the silanyl-substituted material results in direct overlap, while in the hexyl-substituted crystal the pyrene cores overlap more significantly with the sibstituents in adjacent planes. This results in a HOMO-band dispersion of 0.45\,eV primarily along the $a$-axis in the silanyl crystal, which suggests that a high quasi-one-dimensional hole mobility, comparable to what is obtained by vacuum thermal deposition, could be achieved in these solution-processed crystals. However, steady-state photoluminescence measurements reveal evidence of excimer-like photoexcitations, as is found in unsubstituted pyrene crystals, providing photophysical defects that would limit carrier mobility. Substituted pyrene crystals are useful model systems to explore the interplay between electronic transport and photophysics. 

\begin{acknowledgments}
CS acknowledges funding from the Natural Sciences and Engineering Research Council (NSERC), the Canada Foundation for Innovation, and the Canada Research Chairs Program. MC acknowledges funding from NSERC and the R\'eseau qu\'eb\'ecois de calcul de haute performance. We thank Jean-S\'ebastien Poirier for the profilometry measurements. 
\end{acknowledgments}



\section{APPENDIX: FRANCK-CONDON ANALYSIS OF THE ABSORPTION SPECTRA IN SOLUTION}

In this appendix we present a Franck-Condon analysis based on the DFT calculations presented in Section~\ref{Sec:Results} to unravel the vibronic structure of the absorption spectra in Fig.~\ref{fig_crystal}(a). In this model, the intensity distribution of the vibronic replica resulting from a vertical optical transition is given by a Poisson distribution, such that the intensity if the $\jmath$th replica is
\begin{equation}
I_\jmath = \frac{e^{-S}S^{\jmath}}{\jmath!},
\label{eq:FC}
\end{equation}
where $S$ is the Huang-Rhys factor, quantifying a measure of the number of quanta of vibrational energy that are excited within a vibrational normal mode. Within the harmonic approximation, $S$ is defined as
\begin{equation}
S = \frac{1}{2} \frac{k (\Delta Q)^2}{\hbar \Omega},
\label{eq:S}
\end{equation}
where $k$ is the spring constant of the harmonic oscillator of frequency $\Omega$, and $\Delta Q$ is the configurational displacement along a generalized molecular coordinate. Our \emph{ab initio} calculations permit us to obtain the equilibrium configuration in the ground and excited states. The atomic displacement vectors of unsubstituted pyrene following occupation of the LUMO are shown in Fig.~\ref{disp_vect}. However, the Gaussian code does not permit us to calculate the properties of the singlet excited state, which corresponds to an asymmetric linear  combination $1/\sqrt{2}(|\uparrow\downarrow\rangle - |\downarrow\uparrow\rangle)$. We have therefore based this analysis on calculations of the excited triplet state, which can be calculated within a single electronic configuration (such as $|\uparrow\uparrow\rangle$). Based on TDDFT calculations, we find that the first triplet excited state constitutes primarily of the HOMO $\rightarrow$ LUMO transition (its weight in the linear combination is 87\% and 78\% for transitions between the triplet and singlet states, respectively; the (HOMO - 1) $\rightarrow$ (LUMO + 1) contribution is only 3\% in each case). Therefore, we consider that the displacement vectors shown in Fig.~\ref{disp_vect}, based on DFT calculations, can be used to identify the vibrational normal modes that are excited by the optical transition.

We decompose the atomic coordinate displacement vector on the basis of vibrational normal modes. For a molecule containing $N$ atoms and for a displacement vector $\nu$, we define
\begin{equation}
\nu = \sum_{\imath=1}^{3N-6} \alpha_{\imath}\xi_{\imath},
\end{equation}
where $\alpha_{\imath}$ represents an occupation level of the relevant normal mode, and $\xi_{\imath}$ is the vibrational eigenmode. Equation~\ref{eq:S} can then be represented as
\begin{equation}
S_{\imath} = \frac{1}{2} \frac{k \alpha_{\imath}^2}{\hbar \Omega_{\imath}}.
\end{equation}
We then obtain the Huang-Rhys factor for each mode that contributes to atomic displacement as a result of the optical transition.

We have applied this analysis to model the absorption spectrum of Py-Silanyl, and we show the results in Fig.~\ref{FCanal}. The origin of the progression has been adjusted in energy and normalized to match the peak of the experimental spectrum. The energy and relative intensity of the replica are then determined by the DFT calculations. The normal modes contributing to the ground-state absorption with a Huang-Rhys factor greater than 0.10 are tabulated in Table~\ref{FCanalysis}. Each of these modes is associated with a Huang-Rhys factor less than 1, which is consistent with the small Stokes shift of 20\,meV observed in Fig.~\ref{fig_crystal}. These calculated modes are close in energy to those found experimentally by Raman measurements (Fig.~\ref{disp_vect}(b)). Of the modes that contribute the most to the reorganization, modes 142. 158, and 194 correspond to aromatic C---C stretches, and their energies around 180\,meV are typical for this class of modes. Mode 64 is a symmetric breathing mode of the pyrene core.

\bibliography{pyrene}

\newpage
%


 \begin{table}
 \caption{Unit cell parameters for Py-hexyl and Py-silanyl triclinic crystals determined by X-ray diffraction measurements at 77\,K. Both crystals contain only one molecule in the primitive cell.\label{tab_XRD}}
 \begin{ruledtabular}
 \begin{tabular}{c c c}
Parameter & Py-hexyl  & Py-silanyl \\
 \hline
$a$\,({\AA}) &8.4950(3)          &6.3523(6) \\
$b$\,({\AA}) &9.8390(2)          &10.0544(8)\\
$c$\,({\AA}) &12.6215(3)        &15.2949(13)\\
$\alpha$\,($^{\circ}$) &101.514(1)    &108.424(4)\\
$\beta$\,($^{\circ}$) &109.313(1)      &92.212(6)\\
$\gamma$\,($^{\circ}$)  & 96.540(1) & 99.844(6)\\
 \end{tabular}
 \end{ruledtabular}
 \end{table}

  \begin{table}
 \caption{Energy levels of Py-hexyl and Py-silanyl molecules obtained from DFT/B3LYP calculations.\label{energy_molecule}}
 \begin{ruledtabular}
 \begin{tabular}{c c c c}
 & HOMO (eV) & LUMO (eV) & Gap (eV) \\
 \hline
 Py-hexyl & -5.09 & -2.18 & 2.91 \\
Py-silanyl & -5.47 & -2.61 & 2.86 \\
 \end{tabular}
 \end{ruledtabular}
 \end{table}

\begin{table}
 \caption{Normal modes that contribute to the molecular reorganization associated with the ground-state absorption spectrum in isolated Py-silanyl molecules. The mode number refers to its ordering amongst the 240 vibrational modes. The vibrational frequency is $\overline{\nu}$, $E$ is the mode energy, and $S$ is the Huang-Rhys parameter. \label{FCanalysis}}
 \begin{ruledtabular}
 \begin{tabular}{c c c c}
Mode & $\overline{\nu}$ (cm$^{-1}$) & $E$ (meV) & $S$ \\
\hline
64 & 417 & 51.7 & 0.45\\
194 & 1653 & 205.0 & 0.42\\
142 & 1278 & 158.5 & 0.41\\ 
158 & 1380 & 171.1 & 0.19\\ 
52 & 284 & 35.2 & 0.12 
 \end{tabular}
 \end{ruledtabular}
 \end{table}

%



 \begin{figure}
 \caption{Structure of molecules (a) Py-hexyl, which has a pyrene core and hexyl side chains, and (b) Py-silanyl, which has branched chains with a silicon core atom.\label{fig_struct}}
 \end{figure}
 
  \begin{figure}
 \caption{(a) Polarized optical microscopy (POM) view of Py-hexyl crystals drop-casted on a glass slide from an acetone solution (0.3\,mg/mL). Crystalline structure view along the $a$ axis is shown for (b) Py-hexyl and (c) Py-silanyl. Si atoms are shown in yellow.\label{fig_maille}}
 \end{figure}
 
  \begin{figure}
 \caption{View of Py-hexyl crystal along the axis perpendicular to the molecular plane.  Only some molecules have been kept in order to highlight the fact that the side chains from adjacent molecules lie above the pyrene core.\label{49b_stacking}}
 \end{figure}

 \begin{figure}
 \caption{(a) Absorption and emission  spectra of  Py-hexyl (top) and Py-silanyl (bottom) in chloroform at room temperature. (b) Absorption,  emission and photoluminescence excitation (grey) spectra of Py-hexyl (top) and Py-silanyl (bottom) of solution-casted films. The PL and PLE measurements were carried out at room temperature.
 \label{fig_crystal}}
 \end{figure}
 
  \begin{figure}
 \caption{(a) HOMO and (b) LUMO orbitals for pyrene derivative Py-silanyl obtained by DFT calculations.\label{fig_orbitale}}
 \end{figure}

 \begin{figure}
 \caption{Band structure of Py-hexyl (a) and Py-silanyl (b) crystals.  Symbols refer to the following points of the Brillouin zone : $\Gamma = (000$); D = ($\frac{1}{2}0\frac{1}{2}$); B = ($\frac{1}{2}00$); E = ($\frac{1}{2}\frac{1}{2}\frac{1}{2}$); A = ($\frac{1}{2}\frac{1}{2}0$); C = ($0\frac{1}{2}\frac{1}{2}$); F = ($0\frac{1}{2}0$); G = ($00\frac{1}{2}$).  The Fermi energy has been set to 0\,eV.\label{band_structure}}
 \end{figure}

 \begin{figure}
 \caption{(a) Experimental Raman spectrum of Py-sianyl solution in chloroform (black line), and of the chloroform solvent only (grey line). Atomic displacement vectors of unsubstituted pyrene following HOMO--LUMO excitation of triplet states. Carbon atoms are represented as black markers, while hydrogen atoms as red markers. The vector amplitudes are increased by a factor of 20 in order to facilitate their representation. \label{disp_vect}}
 \end{figure}

 \begin{figure}
 \caption{Absorption spectrum of Py-silanyl in chloroform solution (Fig.~\ref{fig_crystal}), with the vibrational replicas calculated as described in this appendix. \label{FCanal}}
 \end{figure}
 
\newpage
\cleardoublepage
\includegraphics[width=\textwidth]{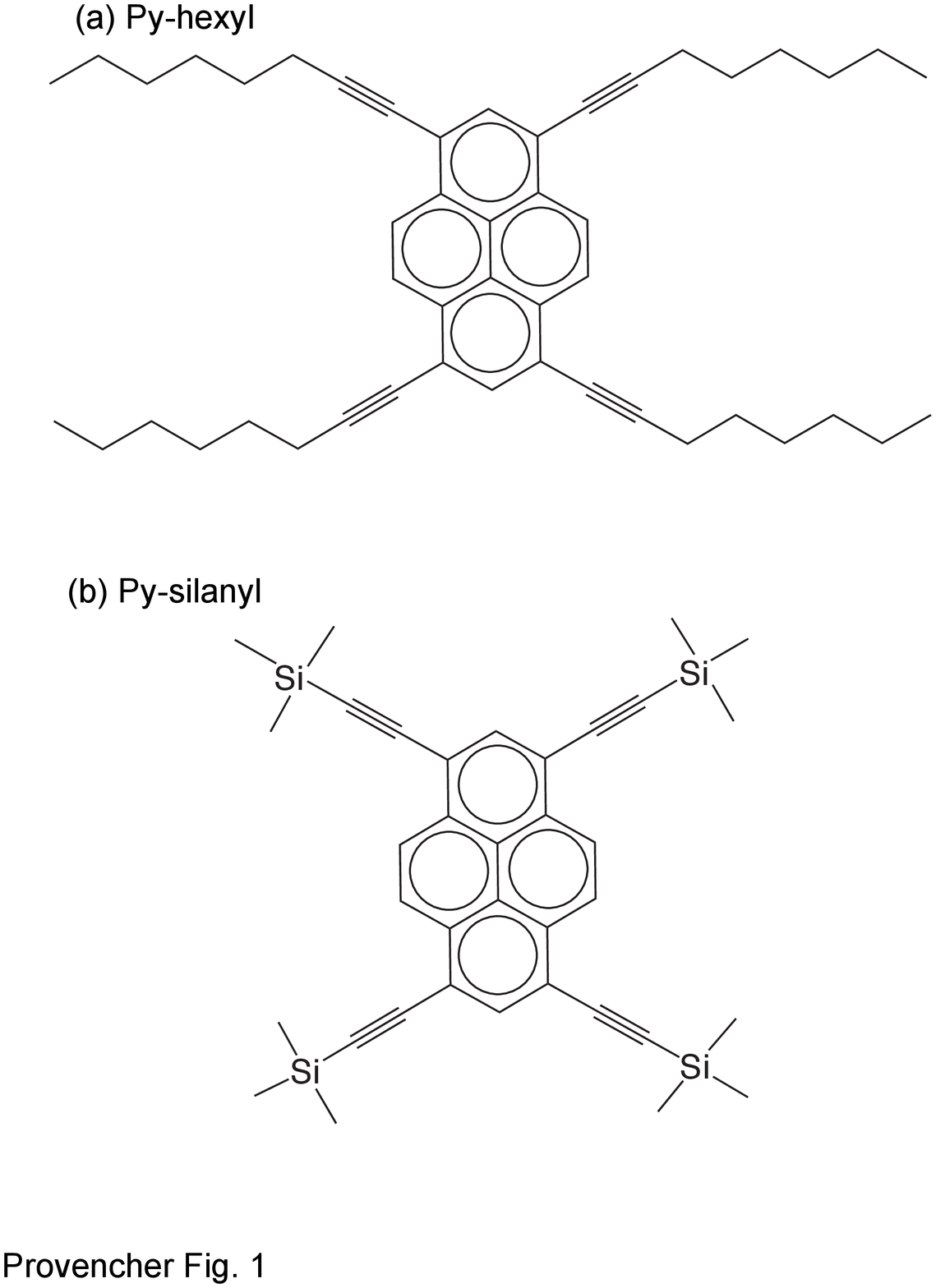}

\newpage
\includegraphics[width=\textwidth]{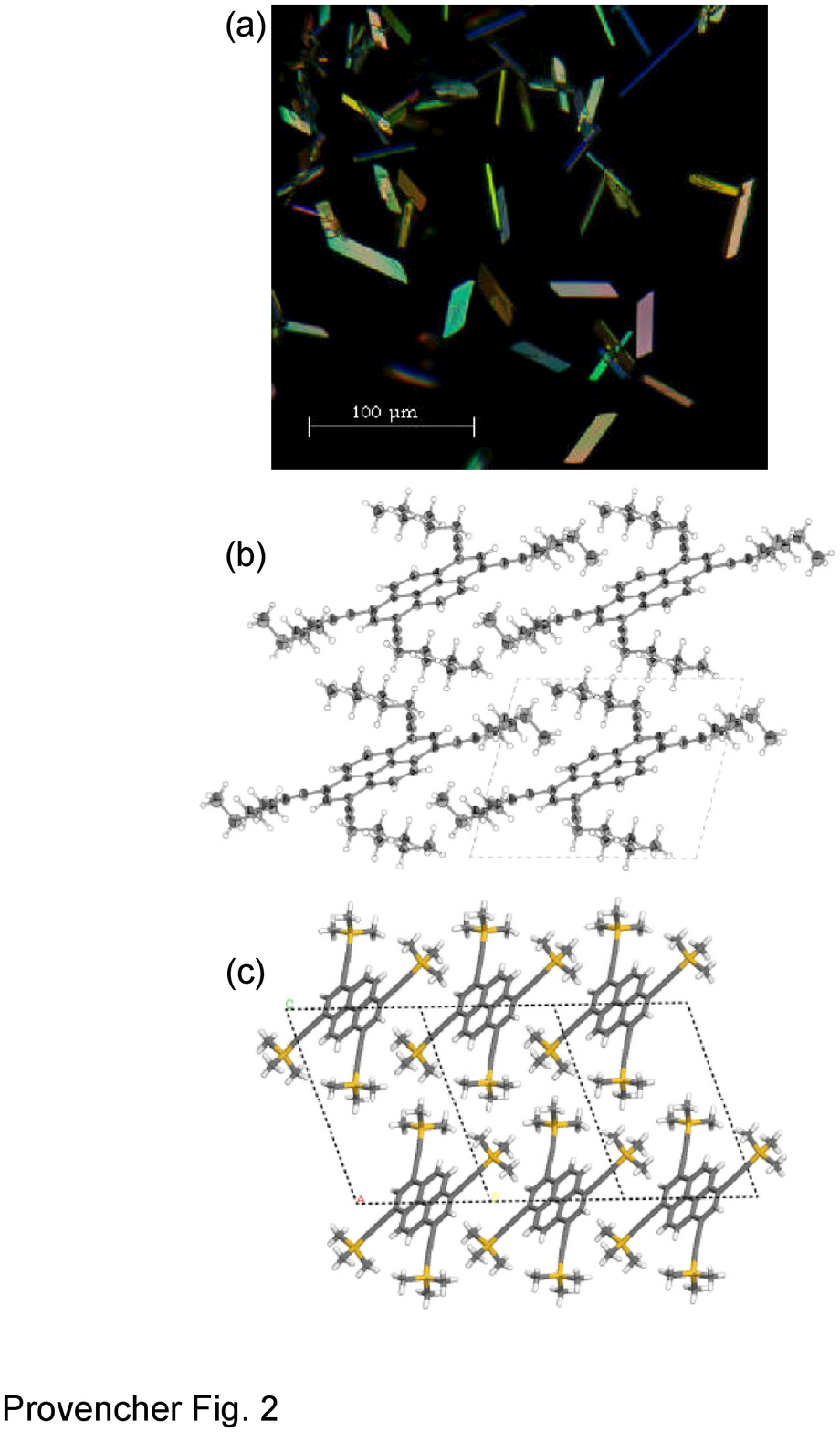}

\newpage
\includegraphics[width=\textwidth]{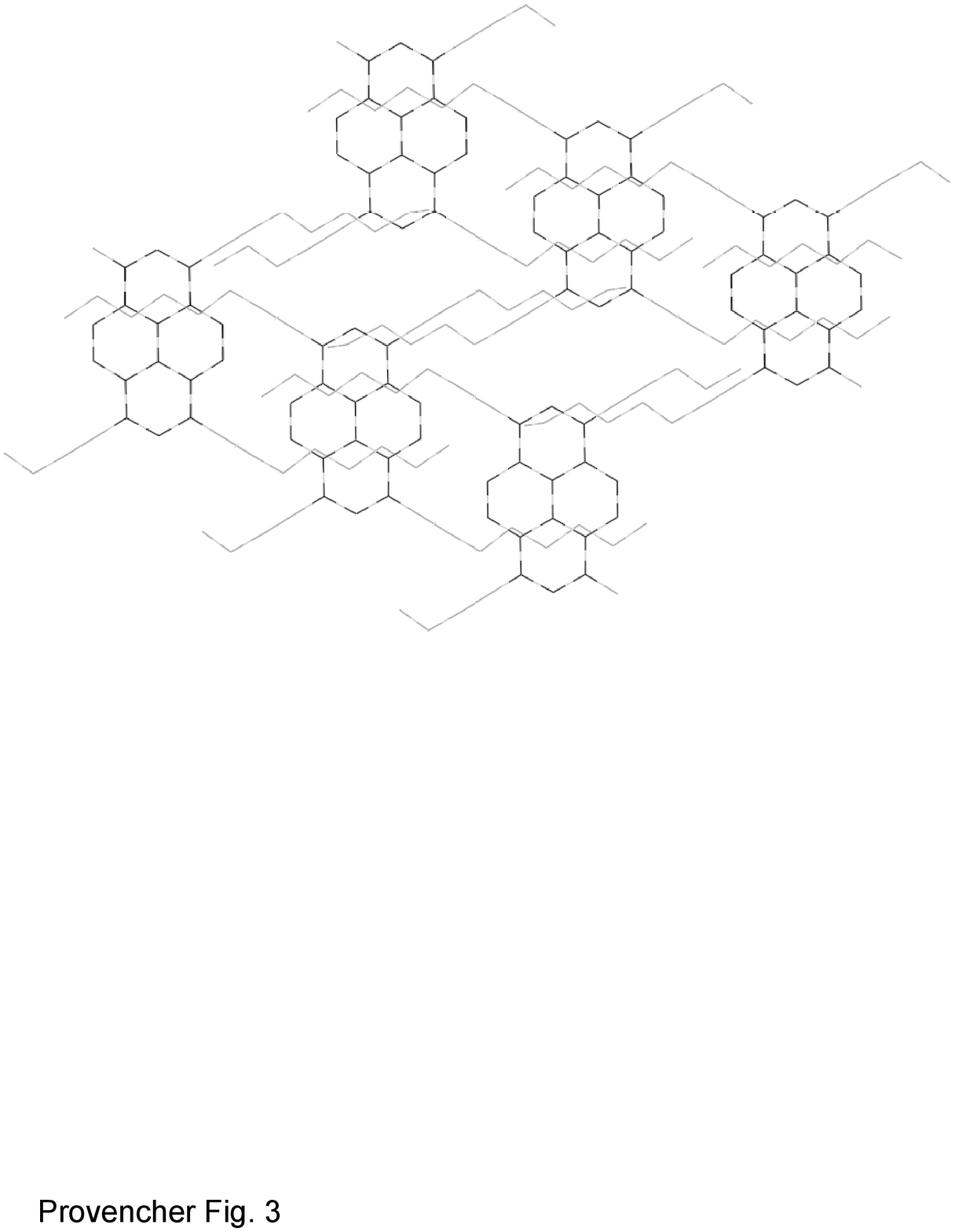}

\newpage
\includegraphics[width=\textwidth]{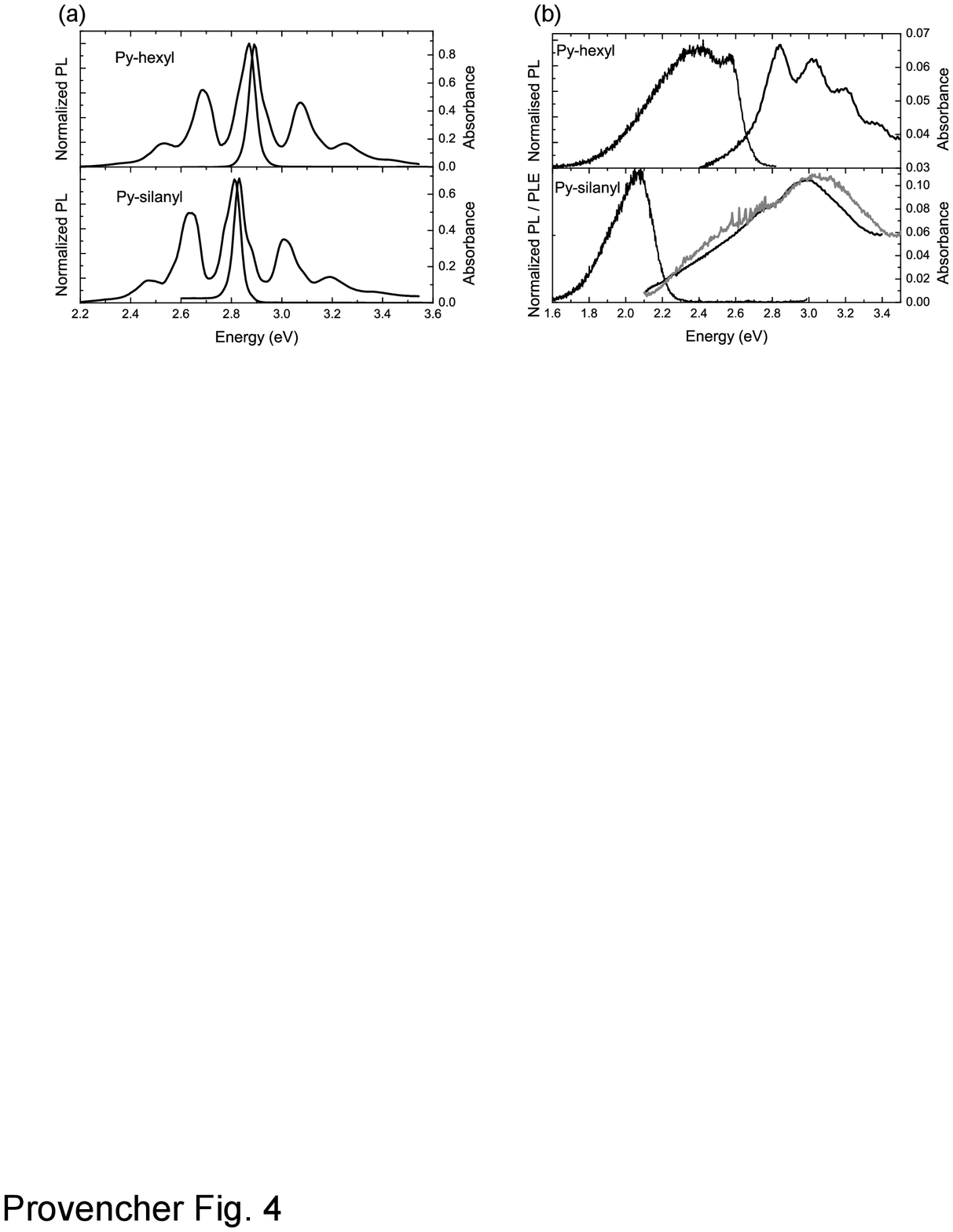}

\newpage
\includegraphics[width=\textwidth]{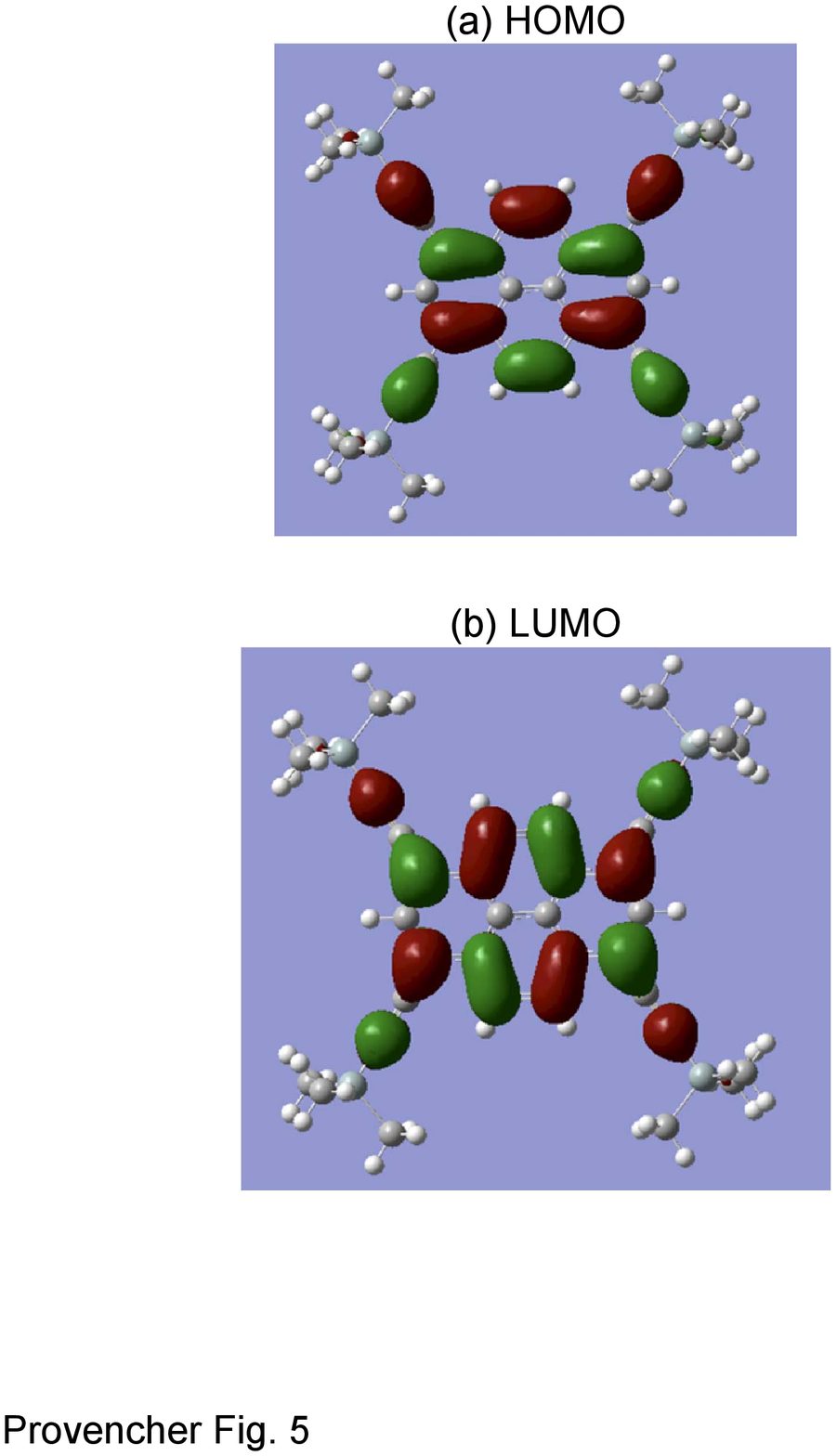}

\newpage
\includegraphics[width=\textwidth]{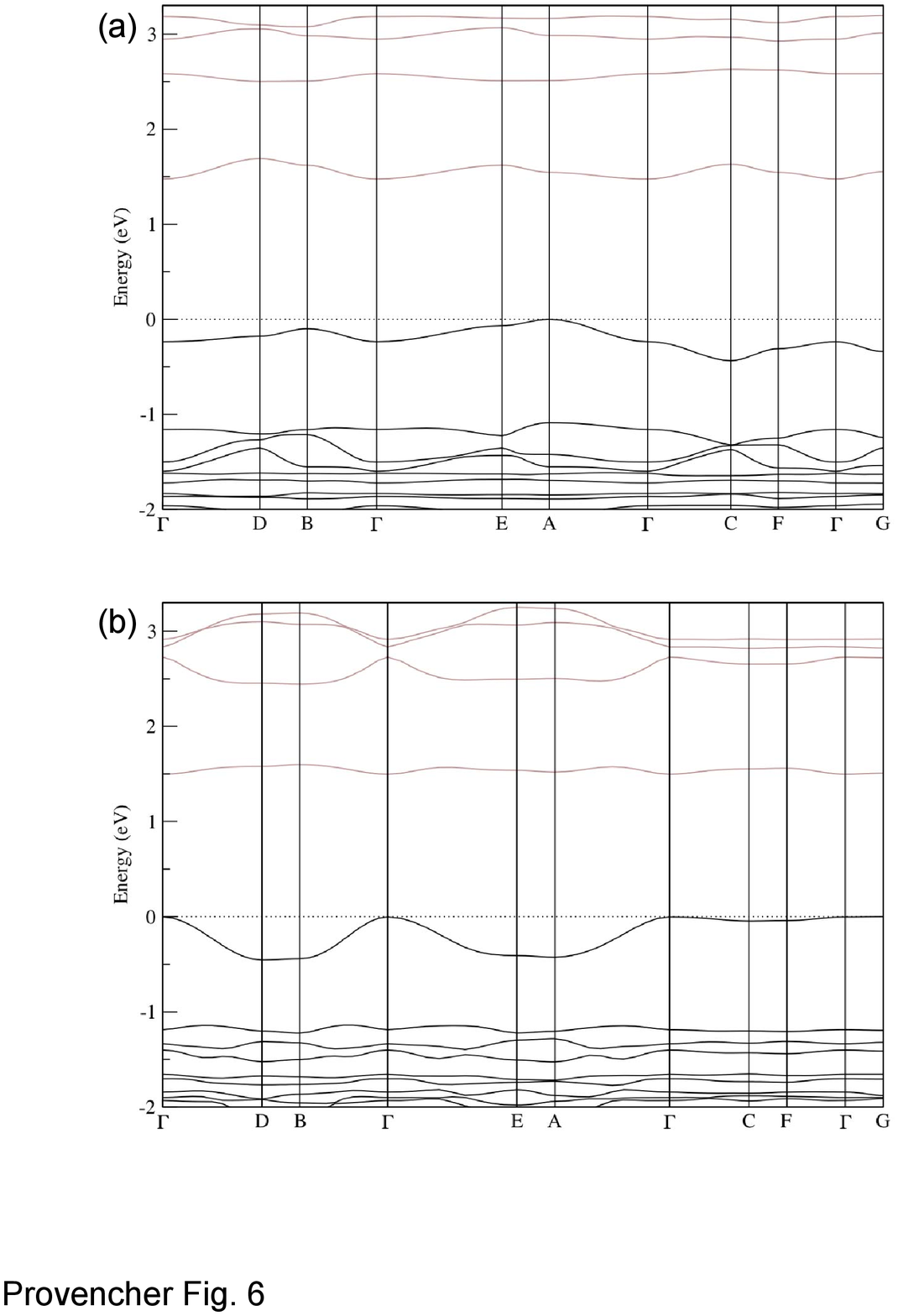}

\newpage
\includegraphics[width=\textwidth]{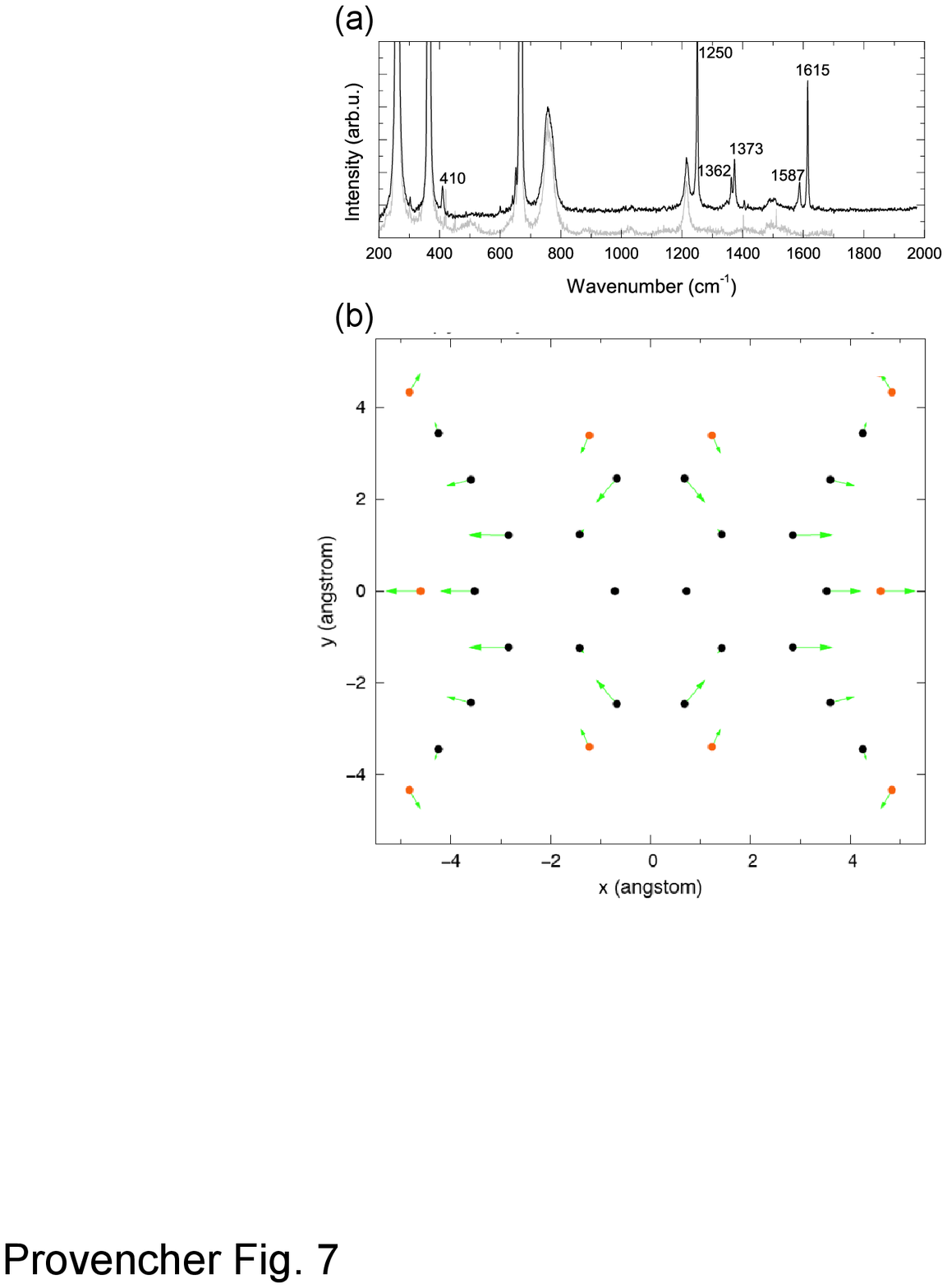}

\newpage
\includegraphics[width=\textwidth]{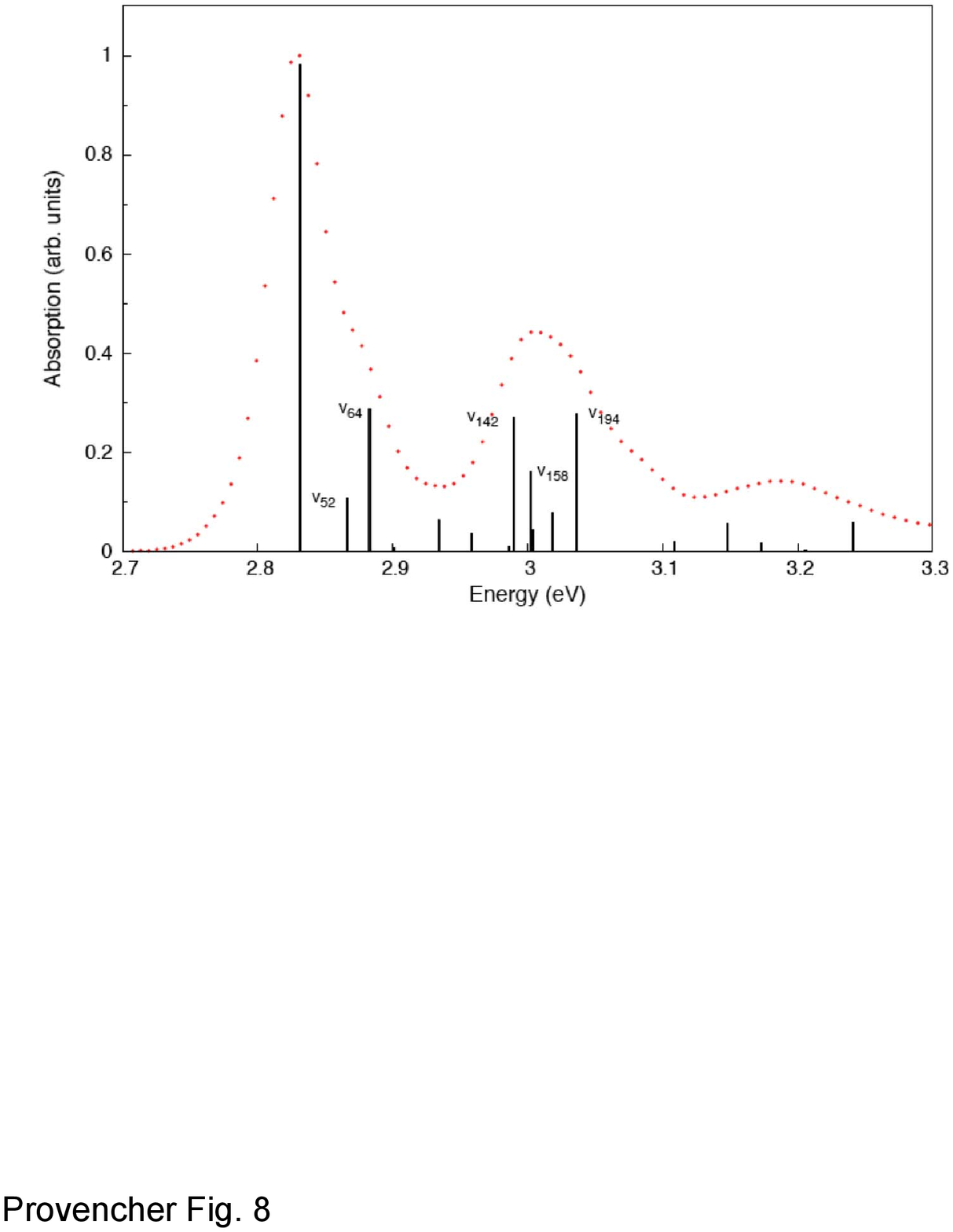}

\end{document}